\documentclass[final,5p,times,twocolumn]{elsarticle}

\usepackage{amssymb}
\usepackage{siunitx}
\journal{Journal of Alloys and Compounds}
\usepackage{subfigure}

\begin{document}

\begin{frontmatter}



\title{A Molecular Study of CaCO$_3$  cluster configurations }


\author[uovg]{\'Angeles Fern\'andez-Gonz\'alez\corref{cor1}\fnref{fn1}}
\ead{mafernan@geol.uniovi.es}
\author[uovf,cinn]{Lucas Fern\'andez-Seivane\fnref{fn2}}
\ead{lucas.fernandez@cinn.es}
\author[uovg]{Manuel Prieto Rubio\corref{cor2}\fnref{fn1,fn3}}
\ead[url]{http://www.geol.uniovi.es}
\author[uovf,cinn,lanc]{Jaime Ferrer\fnref{fn2}}

\cortext[cor1]{Corresponding author}
\cortext[cor2]{Principal corresponding author}
\fntext[fn1]{This is the specimen author footnote.}
\fntext[fn2]{Another author footnote, but a little more longer.}
\fntext[fn3]{Yet another author footnote. Indeed, you can have
any number of author footnotes.}

\address[uovg]{Dept. of Geology. Universidad de Oviedo. C/ J. Arias de Velasco 3007-Oviedo, Spain }
\address[uovf]{Dept. of Physics. Universidad de Oviedo. C/ Calvo Sotelo, 3005-Oviedo, Spain. }
\address[cinn]{Centro de Investigación en Nanomateriales y Nanotecnología (CINN) (Consejo Superior de Investigaciones Científicas - Universidad de Oviedo - Principado de Asturias), Parque Tecnológico de Asturias, 33428 Llanera, Asturias, Spain}
\address[lanc]{Department of Physics, Lancaster University, Lancaster LA1 4YB, UK}

\begin{abstract}

Equilibrium relationships involving solids are based in bulk thermodynamic properties that concern ideal crystals of infinite size. However, real processes towards equilibrium imply nucleation and development of finite molecular-scale entities. The configuration of these early-stage clusters and the estimation of their excess energies with respect to the ideal crystal are key to understand the macroscopic behavior of a given system. Here, starting from the ideal atomic positions in bulk calcite, aragonite, or vaterite, the relaxation of finite clusters of CaCO$_3$ in vacuum has been explored. With the aim of determining the impact of the cluster size on its energy and its geometrical configuration, a series of CaCO$_3$ clusters have been simulated and their lattice energies calculated. The cluster geometry has been fully optimized at constant pressure and its energy has been determined using \texttt{GULP}. A wide variety of clusters ranging from 1 to 2000 formulae has been considered for each starting structure (calcite, aragonite, or vaterite). \texttt{GULP} calculations have been carried out using the pair potentials set derived by Rohl \emph{et al}~\cite{Rohl2003}. In a number of cases the final configuration has been checked using the DFT code \texttt{\texttt{SIESTA}}~\cite{Soler2002,Artacho2008}, and an excellent agreement has been found. Although these simulations do not represent fully realistic scenarios~\cite{Gale2005,Raiteri2010,Raiteri2010a}, some results are relevant from the point of view of the polymorphic precipitation of CaCO$_3$. A thorough analysis of the diffraction patterns of the relaxed clusters has allowed addressing the fundamental question about the critical size that a cluster should have to be considered as truly calcite, aragonite, or vaterite. 
\end{abstract}


\end{frontmatter}


\section{Introduction}
\label{sec:intro}

Equilibrium relationships involving solids are based on bulk thermodynamic properties that concern ideal crystals of infinite size. However, real processes towards equilibrium imply development of finite molecular-scale entities. The configuration of these early-stage clusters and the estimation of their excess energies with respect to the ideal crystal are keys to understanding the macroscopic behaviour of a given system.
As nucleation events are difficult to study experimentally, both because they occur spontaneously and because the nucleus size is very small, atomistic simulations are a suitable tool for understanding the early stages of crystallisation. Here, starting from the ideal atomic positions in calcite and aragonite, the relaxation in vacuum of finite clusters of CaCO$_3$ is explored.
Nucleation and growth of calcium carbonate phases constitute a very important subject of research in a wide variety of fields. A complete study of CaCO$_3$ should include many different aspects: size and shape of the critical nuclei under diverse conditions, possibility of nucleation from precursor phases, nucleus energy and nucleus surface energy, relationship nucleus-substrate in heterogeneous nucleation… We present a preliminary study of nucleation of calcium carbonate where nuclei are considered to be isolated from any previous phase or substrate. Even when this situation does in no way represent realistic conditions, it can be a helpful first approach to more complex studies.
In ambient conditions of pressure and temperature, calcite is the most stable phase of CaCO$_3$. However, if the Ostwald-Lussac law of phases were strictly true, the crystallisation of calcite would begin with nucleation of an amorphous material, which would then transform into vaterite, then aragonite, and finally calcite. Actually, numerous experimental works have shown the existence of precursor phases on calcite crystallisation.

\section{Calculations: method and procedure}
With the aim of determining the influence of the cluster size on its energy and on its geometrical configuration, a series of virtual CaCO$_3$ clusters was built and their lattice energy calculated. For each simulation run the starting configuration was an isolated spherical and electrically neutral cluster of calcium carbonate (calcite or aragonite) containing a definite number of CaCO$_3$ formulae. The cluster geometry was fully optimized at constant pressure and its energy was determined using the \texttt{GULP} program (\cite{Gale1997,Gale2003}). A wide variety of clusters ranging from 1 to 2000 formulae was considered for both calcite and aragonite initial configurations
For these calculations atomistic simulation techniques based on the Born model of solids were employed. In this method, the atoms are considered to be charged balls, which are free to interact with each other. These interactions are long-range electrostatic forces and short-range forces which can be described using simple analytical functions. The components of the short-range forces include both the repulsions and the van der Waals attractions between neighbouring electron charge clouds. \texttt{GULP} calculations were carried out using the pair potentials set derived by Rohl \emph{et al}~\cite{Rohl2003}. This set of potentials includes the polarizability of the carbonate ions in the system via the shell model of Dick and Overhauser~\cite{Dick1958}. The oxygen ions are described by a shell of zero mass representing the electronic charge cloud connected to a core containing all the mass of the ion. The total charge of the ion is the sum of the charges of core and shell. The position of the core represents the position of the ion in the crystal lattice. The shell and core are connected by a harmonic spring with a force constant. All intramolecular potentials are defined to act between cores while intermolecular interactions act on shells, where applicable. Furthermore, all Coulomb interactions are excluded within molecules. Although the shell model is relatively simple, it has been very successful in modelling a variety of properties. This set of potentials is transferable between the different CaCO$_3$ polymorphs and it was shown to be excellent at simulating surfaces and environments with low coordination (\cite{Gale2003,Austen2005,Magdans2006,VINOGRAD2007}) .The parameters of the force field are given in \ref{tab:Table1}.
\begin{table*}[t]
\scriptsize
\centering
\begin{tabular}{c c c c c c c c c c c c c c c}
\hline
Atom 1 & Atom 2 & Atom 3  & A  & B & C & D$_e$ & a$_0$ & r$_0$ & k$_{cs}$ & k$_\theta$ & $\theta_0$ & k$_2$ & k$_4$ & Cutoff \\
       &        &         &(eV)&(\AA)  & (eV \AA$^6$) &eV &(\AA$^{-1}$) &(\AA) &(eV\AA$^{-2}$) &(eVrad$^{-2}$)& ({}$^\circ$) &(eV\AA$^{-2}$) &(eV\AA$^{-4}$) &(\AA) \\ \hline
\multicolumn{15}{ c }{Buckingham potentials: $E(r) = A \exp(-r/B) - C/r^6$ } \\ \hline
O  core	& O core	 &    &	4030.3	& 0.245497	& 0.0	&	&	&	&	&	&	&	&	&             Intra/2.5 \\
O  shell	& O shell & 	&	64242.454	& 0.198913	& 21.8436	&	&	&	&	&	&		&	&  &   Inter/15.0\\
Ca core	& O shell & 	&	2154.0	& 0.289118	& 0.0	&	&	&	&	&	&	&	&	&   10 \\
Ca core	& C core	 & 	&	1.2x108	& 0.120	& 0.0	&	&	&	&	&	&	&	&	   &   10 \\ \hline
\multicolumn{15}{ c }{Morse potential: E(r) = De({1-exp[-a0(r-r0)]}2-1) } \\ \hline
C core	& O core	 &	 &		& & & 5.0	& 2.5228 &	1.1982	&	&	&	&	&	& Bonded \\ \hline
\multicolumn{15}{ c }{Spring potential: E(r) = 0.5 kcsr2}  \\ \hline 
O core	& O shell	 &			&	&	&	&	& & & 5.740087	&	&	&	&	& 0.8 \\ \hline 
\multicolumn{15}{ c }{Three-body potential (pivot = atom1): E(r) = 0.5k ( - 0)2 }\\ \hline 
C core	& O core & O core	&	&	&	&	& & & & 1.7998	& 120.0	&	&	& Bonded \\ \hline 
\multicolumn{15}{ c }{Out of plane potential (pivot = atom 1): E(d) = k2d2+k4d4 }  \\ \hline 
C core	& O core	&	&	&	&	&	&	&	&	&	&		& 8.6892	& 360.0 &	Bonded \\ \hline
\end{tabular}
\caption{Parameters of the force field. Carbonate anions are defined as molecules, within which Coulomb interactions are excluded and C/O atoms are deemed to be bonded. All atoms are represented by a core, except O atoms, which also have a shell. The charges are: Ca core = +2.0, C core = +1.343539, O core = +1.018487, and O shell = -2.133.}
\label{tab:Table1}
\end{table*}

In order to ensure the validity of this force field in the simulation of very small clusters, the starting configurations with 10 or less formulae unit were also optimised and their energies calculated with the DFT-based code \texttt{SIESTA} (\cite{Soler2002,Artacho2008}). For these calculations the pseudopotentials for Ca, C and O by Archer (2004) were used and a double Z-polarized basis set was considered~\cite{Archer2004}. The energies and final relaxed configurations were verified to be consistent with the GULP results.
The final atomic coordinates of the relaxed clusters were compared with the crystal structure of calcite or aragonite. With this aim, the X-ray diffraction patterns of all the final relaxed clusters were calculated with the Debye formula and compared with the corresponding aragonite or calcite pattern.

\begin{figure*}[t!]
\centering
\subfigure{\label{fig:fig1a}\includegraphics[width=0.95\columnwidth ]{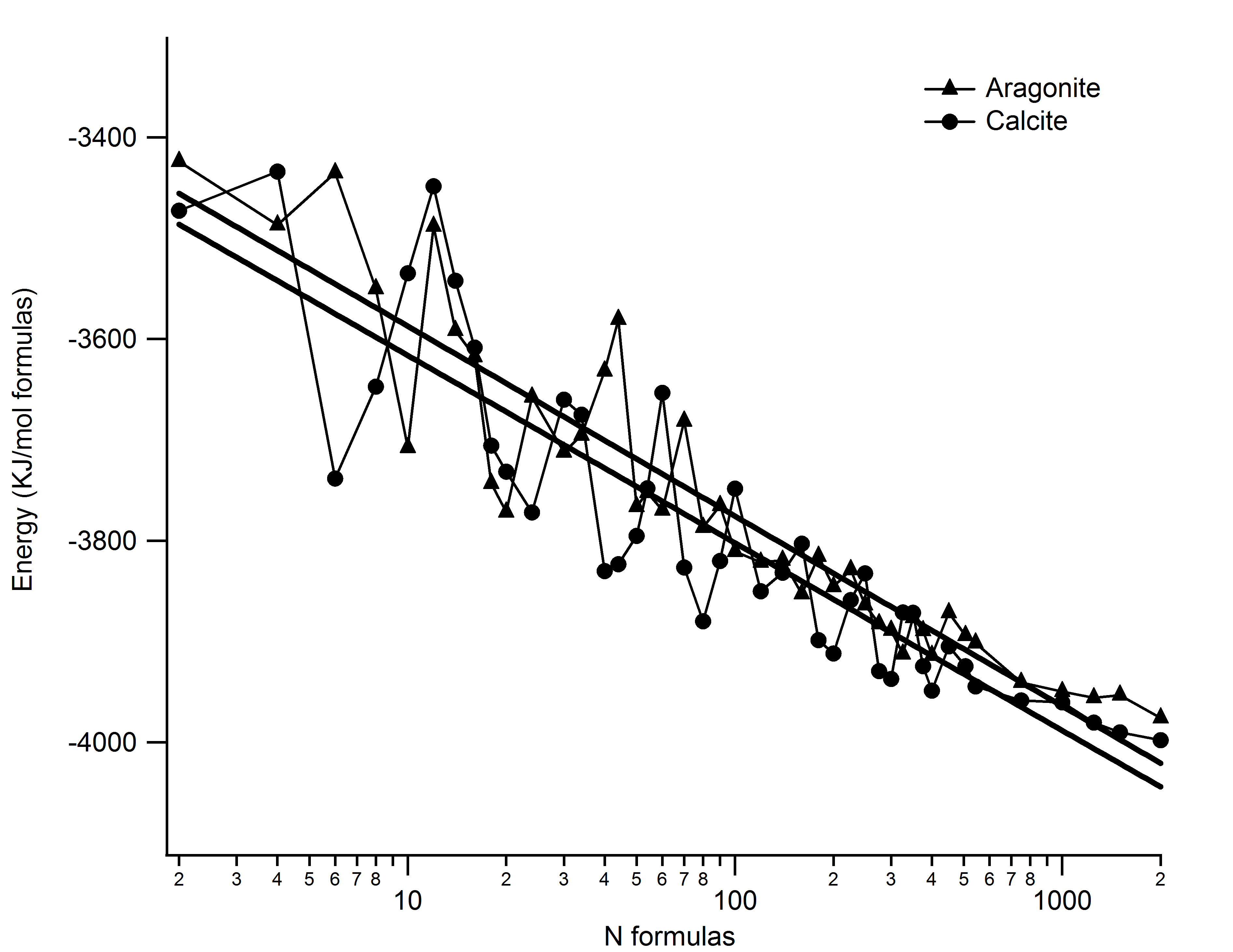}}
\subfigure{\label{fig:fig1b}\includegraphics[width=0.95\columnwidth ]{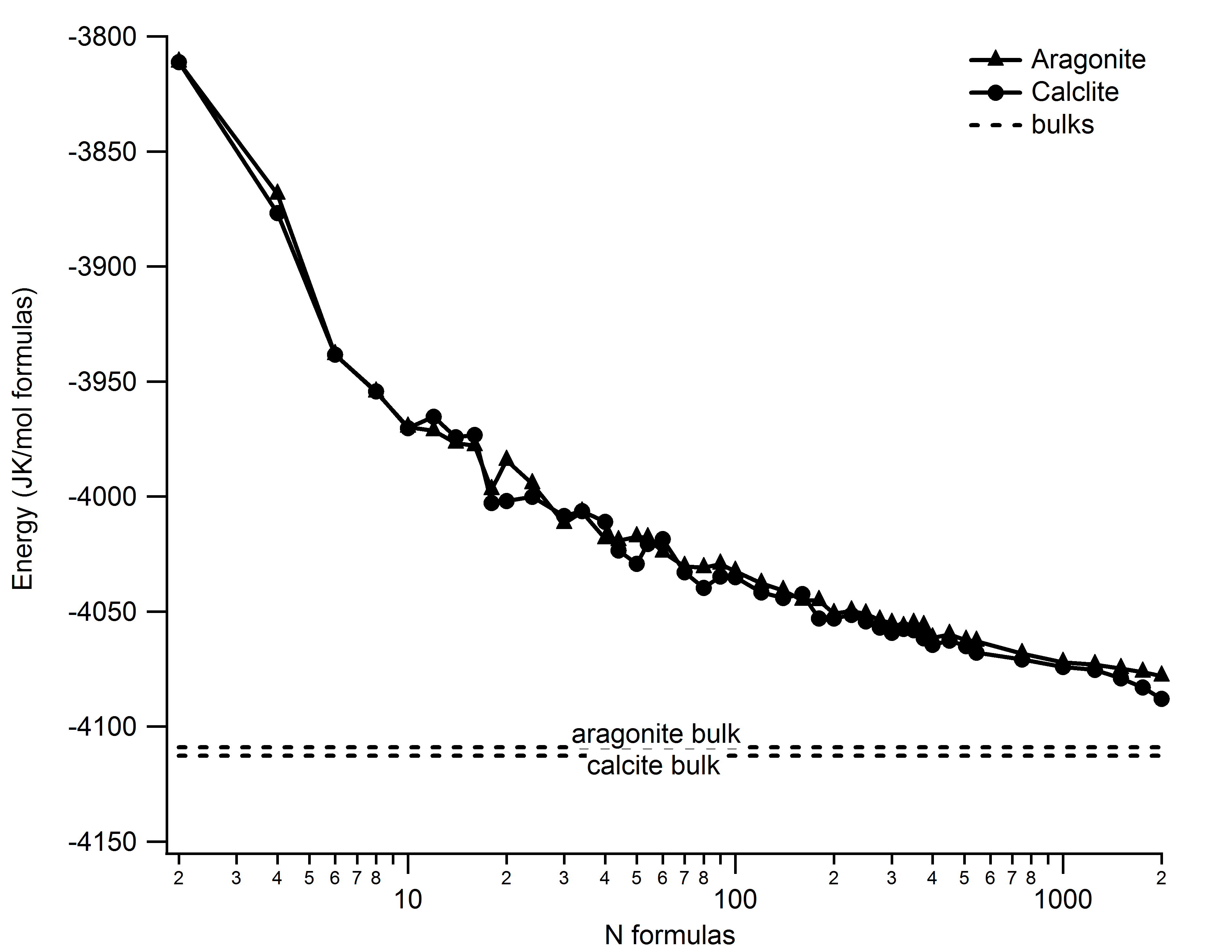}}
\caption{Energy of un-relaxed [\subref{fig:fig1a}] and relaxed [\subref{fig:fig1b}] clusters. The horizontal dashed lines in \subref{fig:fig1b}] represent the lattice energies of the corresponding bulks, calcite and aragonite. The logarithmic scale chosen for the X axis allows to observe the big differences of energy in small clusters.}
\label{fig:1EnerUnrel}
\end{figure*}

\begin{figure}[th]
\centering
\includegraphics[width=\columnwidth ]{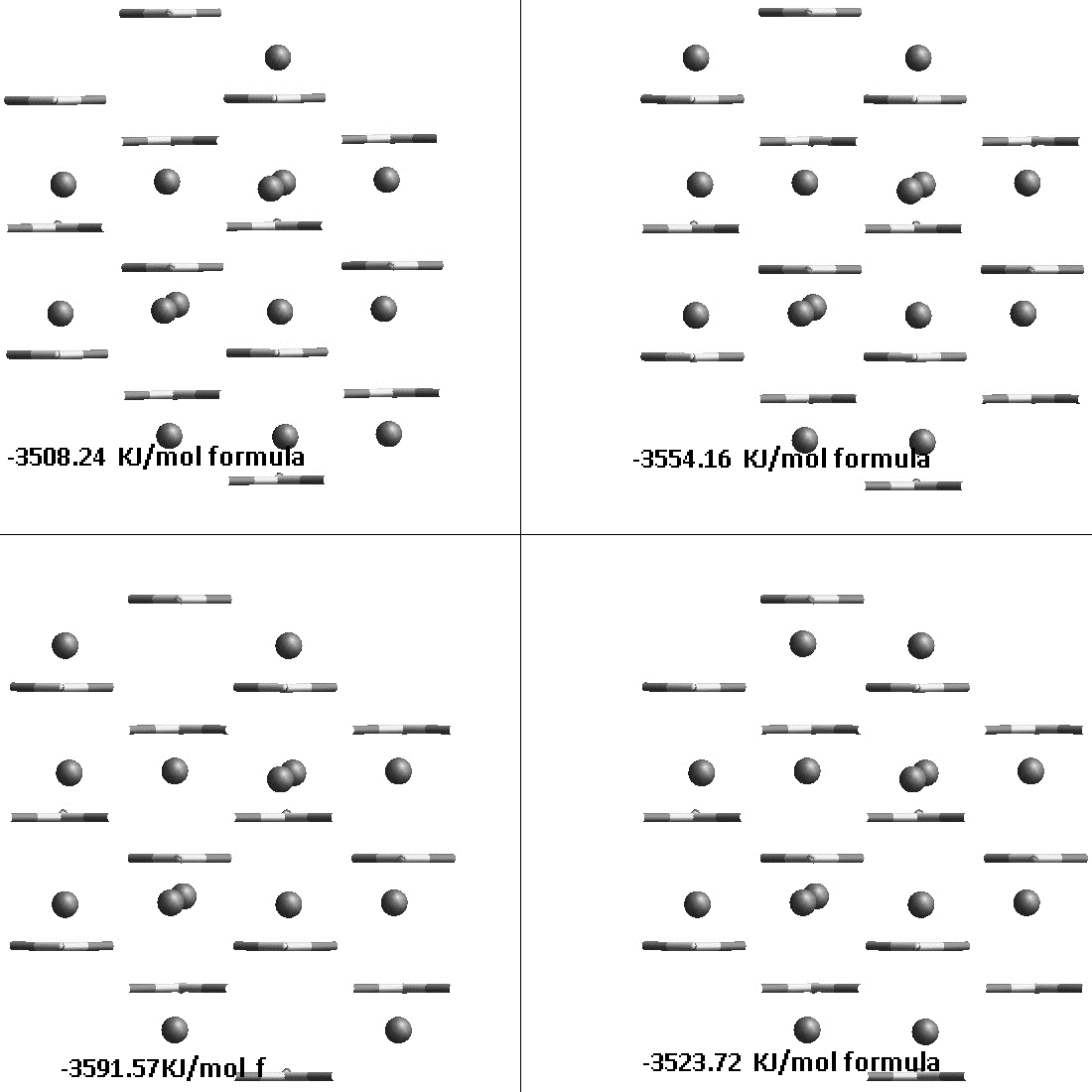}
\caption{Example of four different configuration of aragonite clusters with 14 formulae. All of them have the same radius and chemical content, but their energies are significantly different.}
\label{fig:2Configs}
\end{figure}

\section{Results}
The calculated total energies (in \kilo\joule \per \mole formula) of calcite and aragonite un-relaxed clusters from 1 to 2000 CaCO$_3$ formulae are represented in \ref{fig:fig1a}. In some cases (mainly for small clusters with less than 30 formula units), there are diverse initial possible configurations with the same number of formulae and similar radius, whose calculated energies are considerably different. For example, three different aragonite clusters with 14 formula units are represented in \ref{fig:2Configs}. As can be seen, their calculated energies per mol are significantly different. In such cases, the energies of every possible configuration were calculated, and the one with lower energy was plotted in \ref{fig:fig1a}. The general trend is similar in both calcite and aragonite cases: there is a decrease in energy per formula with the increasing number of formulae in clusters. This increase is most important in clusters with a small number of CaCO$_3$ formulae. As is shown in the plots, data can roughly be fitted to logarithmic functions of the number of formulae, but the energy decrease is not continuous and there are specific cluster sizes that are particularly stable. In these cases the increment of formulae (in a small interval) means an increase in the cluster energy. For example, the un-relaxed calcite cluster with  24 formulae is more stable than clusters with 30 and 34 formulae; or the un-relaxed aragonite cluster with 10 formulae is more stable than clusters with 12, 14 and 16 formulae. These deviations from the general tendency are most important in clusters with a small number of formulae.

The calculated total energy of the relaxed clusters (in \kilo\joule \per \mole formula) is represented in \ref{fig:fig1b}. The horizontal dashed lines below the curves represent the energies of calcite and aragonite bulks calculated with the same set of potentials. As in un-relaxed clusters, a decrease is energy per formula with the increasing number of formulae is observed. Here, the divergence of the data from the general tendency is significantly lower than in the case of un-relaxed clusters. Explored clusters do not reach energies as low as calcite or aragonite bulks. A 2000 formulae relaxed cluster with calcite starting configuration is about 25~\kilo\joule \per \mole more energetic than the corresponding calcite crystal, and a 2000 formulae relaxed cluster with aragonite starting configuration is about 30~\kilo\joule \per \mole more energetic than the aragonite bulk.

\section{Discussion}
For clarity in discussion, clusters will be classified in three groups. The first group will include small clusters, with less than 12 formulae. In this case, all the structural units (Ca or CO$_3$) are in the cluster surface and all the coordination polyhedra of Ca are incomplete. Clusters with intermediate size between 14 and 160 CaCO$_3$ formulae will be considered in a second group. The un-relaxed clusters belonging to this second group have a considerable number of structural units in their surface, but some coordination polyhedra of Ca are complete. The third group will include clusters from 170 to 2000 CaCO$_3$ formulae. In un-relaxed clusters of this group most of Ca complete their coordination polyhedron.
\subsection{Small clusters}
\begin{figure}
\centering
\includegraphics[width=\columnwidth ]{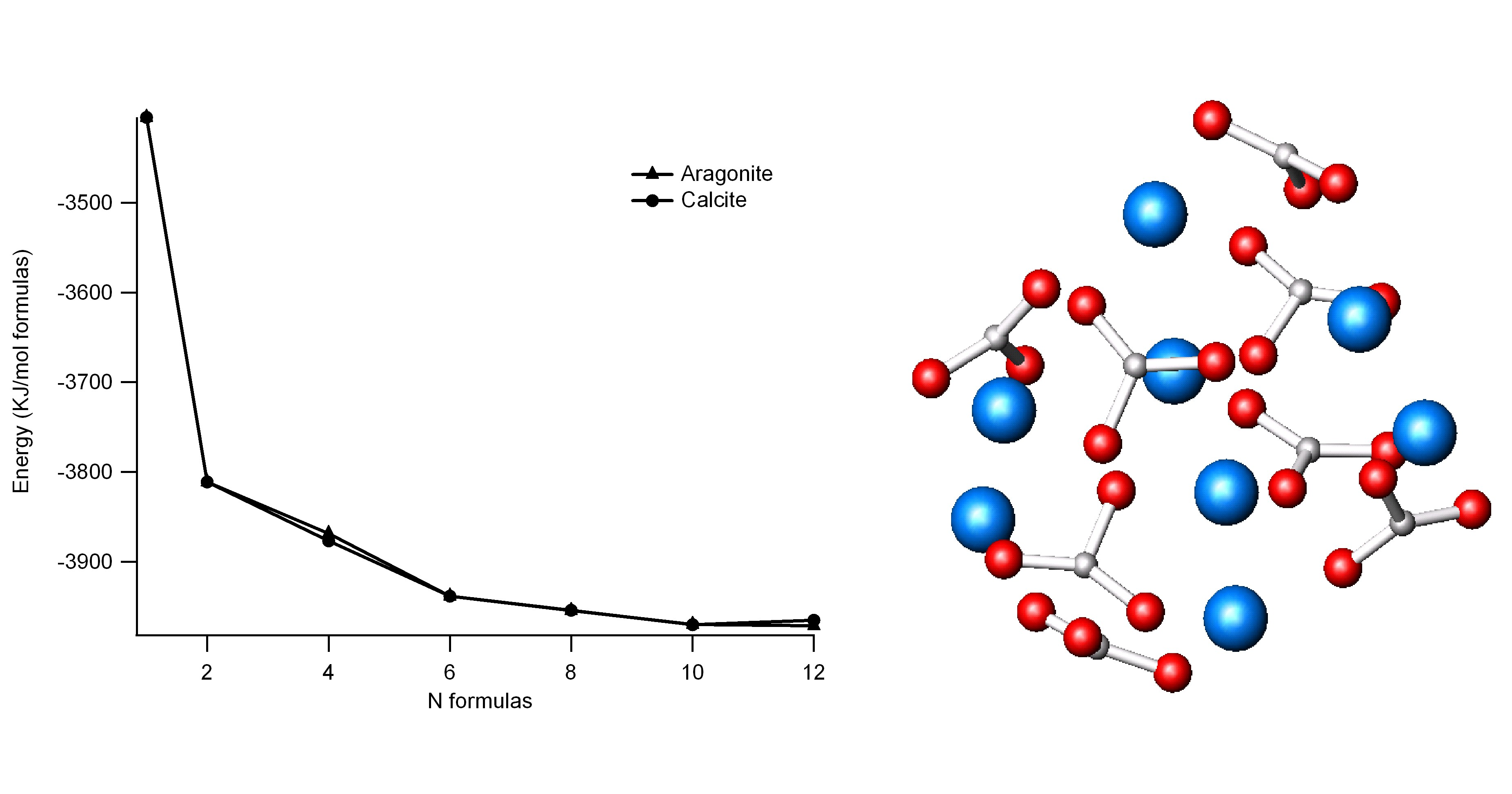}
\caption{Energy of small relaxed clusters (a). Example of relaxed small cluster with 8 formulae (b).}
\label{fig:3Enrgmorphrel8form}
\end{figure}
The energies of small un-relaxed and relaxed CaCO$_3$ clusters whose initial structure is calcite or aragonite are compared in \ref{fig:3Enrgmorphrel8form}A. Small clusters of calcite and aragonite relax to the same or to equivalent relaxed configurations with the same energy.
As a representative example, a relaxed cluster with 8 CaCO$_3$ is reresented in \ref{fig:3Enrgmorphrel8form}B. No order between calcium and carbonate can be found in relaxed small clusters. As expected, the simulated X-ray diffraction patterns of small clusters do not show any significant reflection. These patterns are included in \ref{fig:6Enrgmorphrel8form}.

\subsection{Intermediate size clusters}
\begin{figure}
\centering
\includegraphics[width=\columnwidth ]{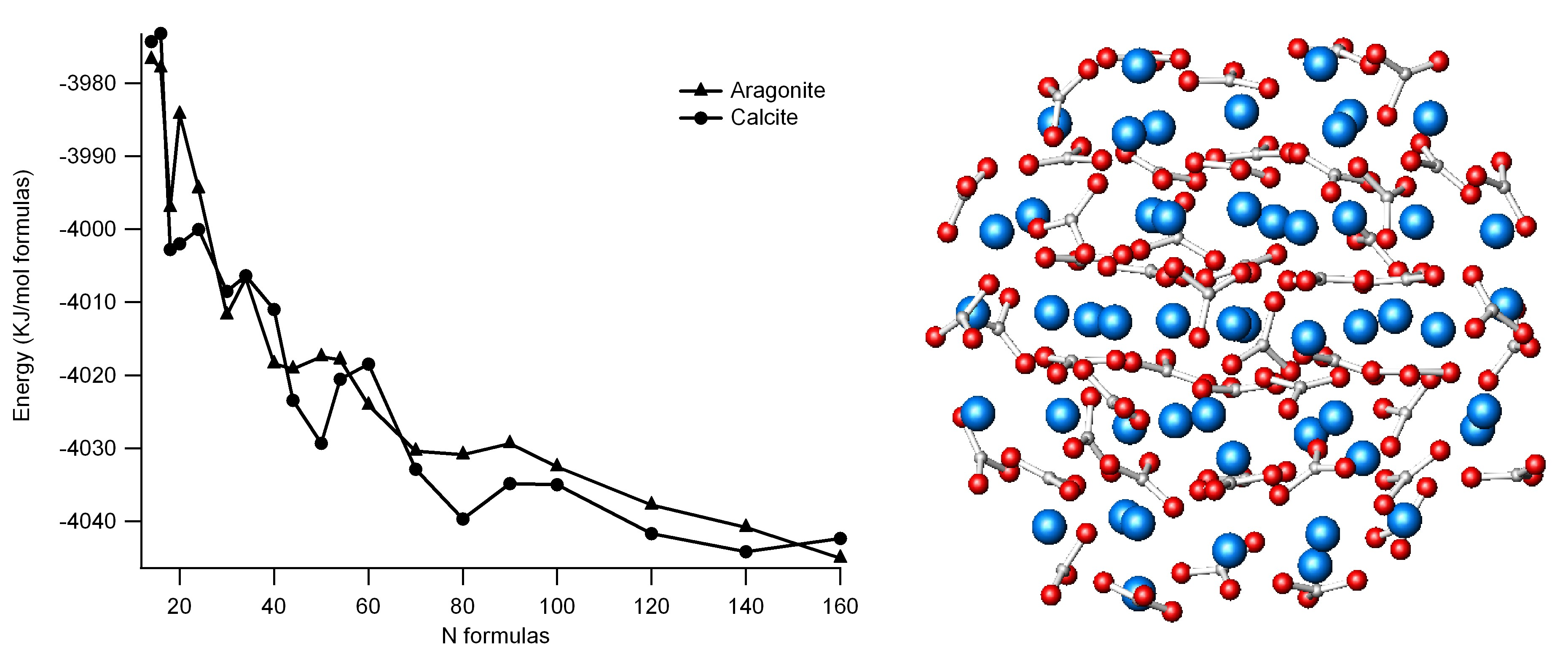}
\caption{Energy of intermediate size relaxed clusters (a). Example of relaxed intermediate cluster with 50 formulae (b)}
\label{fig:4Enrgmorphrel8form}
\end{figure}

The energies of intermediate relaxed CaCO$_3$ clusters are compared in \ref{fig:4Enrgmorphrel8form}A. Energies and configurations of relaxed intermediate clusters are clearly different depending on their initial configuration. In some cases clusters relaxed from the aragonite configuration are more stable and in other cases clusters derived from the calcite configuration are less energetic.
A representative example of relaxed intermediate cluster is shown in \ref{fig:4Enrgmorphrel8form}B. Relaxed intermediate clusters of calcite and aragonite are partially ordered. The rows of alternating Ca and CO$_3$, characteristic of calcite and aragonite structures, can be observed in the inner part of the relaxed clusters. Moreover, in this inner part, the relative orientation of the CO$_3$ groups is almost entirely preserved and in general, the Ca coordination polyhedra can still be identified although they are distorted. However, the structural units in the cluster surface relax to clearly different positions. Carbonate triangular groups tend to be positioned tangent to the cluster surface, and their O atoms try to complete the coordination polyhedra of the superficial Ca atoms.
Despite the relative disorder in relaxed clusters, the simulated X-ray diffraction patters show peaks that can be identified as important reflections of the initial structure calcite or aragonite. Simulated X-ray diffraction patterns of intermediate clusters are included in \ref{fig:6Enrgmorphrel8form}.

\begin{figure}
\centering
\includegraphics[width=\columnwidth ]{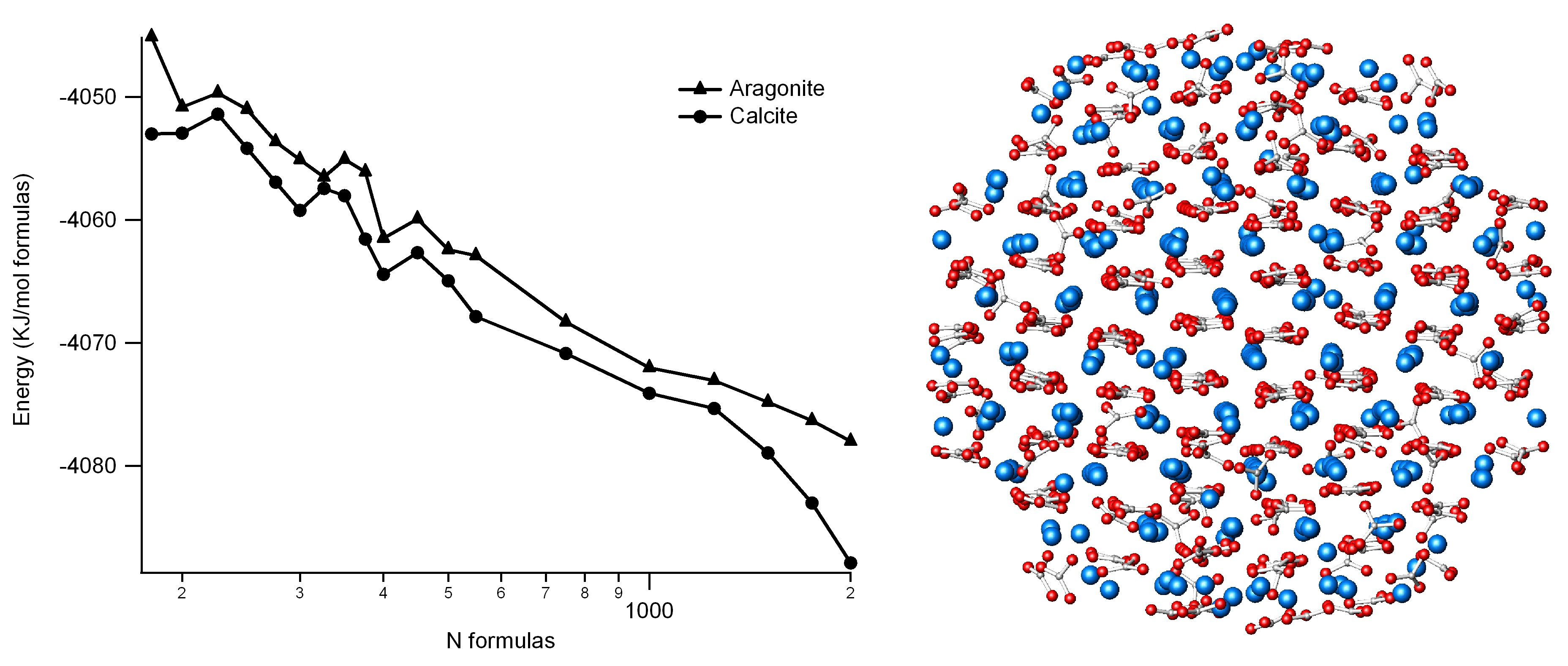}
\caption{Energy of large relaxed clusters (a). Example of relaxed intermediate cluster with 300 formulae (b)}
\label{fig:5Enrgmorphrel8form}
\end{figure}

\subsection{Large clusters}
\begin{figure}
\centering
\includegraphics[width=\columnwidth ]{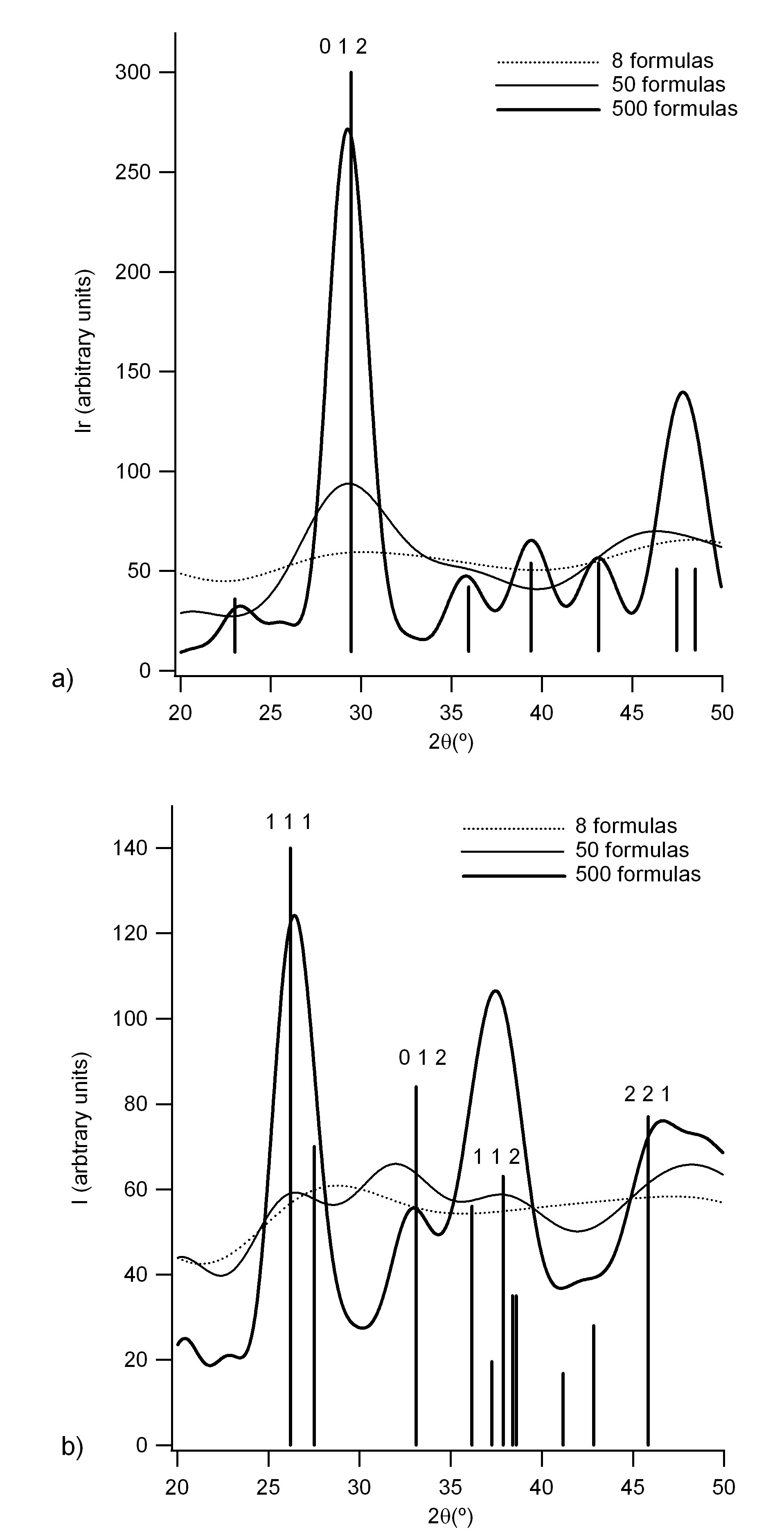}
\caption{Simulated diffraction patterns of representative caclite-like (a) and aragonite-like relaxed clusters (b). The main reflections of aragonite or calcite are shown as bars}
\label{fig:6Enrgmorphrel8form}
\end{figure}

The energies of large relaxed clusters of CaCO$_3$ are shown in \ref{fig:5Enrgmorphrel8form}A. In large clusters, the final energy of relaxed configurations depends on the initial configuration. As can be seen, all relaxed clusters explored in this range have lower energy if the initial configuration is calcite-like.
An example of a large relaxed cluster is shown in \ref{fig:5Enrgmorphrel8form}B. Relaxed clusters belonging to this group show the calcite or aragonite structure of the corresponding initial un-relaxed cluster. Except in their surface, the inter-atomic distances and coordination polyhedra are essentially the same as those of calcite or aragonite bulks. As in intermediate clusters, the main effect on relaxation affects to the planar CO$_3$ groups in the cluster surfaces, which move and rotate to be positioned tangent to the surface.
As can be seen in \ref{fig:6Enrgmorphrel8form}, the simulated X-ray diffraction patterns of large clusters are very similar to the corresponding aragonite or calcite diffractograms.

\section{Conclusions}

Although the simulations carried out in this work do not represent any fully realistic scenarios, we can find some relevant conclusions from the point of view of the polymorphic precipitation of CaCO$_3$. 
Small stable clusters of calcium carbonate that can be interpreted as nuclei in the first stages of nucleation, do not present calcite or aragonite structure. For CaCO$_3$ entities below 12 formulae, the amorphous configuration is more stable than any other ordered phase considered. Even though calcite is the more stable polymorph of calcium carbonate at room temperature, some aragonite clusters in the range of 12-160 CaCO$_3$ formulae are more stable tan the corresponding calcite clusters.
One of the reasons of the existence of precursor amorphous or aragonite phases in the crystallization of calcite can be linked with these conclusions. Nevertheless, this study shows that the classical theory of nucleation is insufficient when an atomistic approach to nucleation is required.

\bibliographystyle{model1a-num-names}
\bibliography{elsarticle-template-harv}





\end{document}